\PassOptionsToPackage{unicode}{hyperref}
\PassOptionsToPackage{naturalnames}{hyperref}

\documentclass[reprint, secnumarabic, amssymb, nofootinbib,tightenlines, nobibnotes, aps, prd]{revtex4-2}

\usepackage{graphicx}% Include figure files
\usepackage{dcolumn}% Align table columns on decimal point
\usepackage{bm}% bold math
\usepackage{multirow}
\usepackage{amssymb,mathrsfs,amsmath}
\usepackage{amsfonts}
\usepackage{color,xcolor}
\usepackage{subfigure}
\usepackage{amsmath}
\usepackage{slashed}
\usepackage{soul}
\def\linkcolor{cyan!70!black}

\usepackage[
colorlinks=true
,urlcolor=\linkcolor
,anchorcolor=\linkcolor
,citecolor=\linkcolor
,filecolor=\linkcolor
,linkcolor=\linkcolor
,menucolor=\linkcolor
,linktocpage=true
,pdfproducer=medialab
,pdfa=true
]{hyperref}
\newcommand{\eq}[1]{\begin{align}#1\end{align}}

\begin{document}
\preprint{HEP-IFUNAM, 24-3}

\title{
Lepton number violating and conserving heavy baryon four-body decays in the presence of two almost degenerate heavy neutrinos
}% Force line breaks with \\

\author{Fabiola Fortuna}%
\email{fabyfortuna@fisica.unam.mx}\affiliation{Instituto de F\'{\i}sica,  Universidad Nacional Aut\'onoma de M\'exico, Apartado Postal 20-364, Ciudad de M\'exico 01000, M\'exico}%

\author{Gerardo Hern\'andez-Tom\'e}
\email{gerardo_hernandez@uaeh.edu.mx}
\affiliation{Área Académica de Matemáticas y Física, Universidad Autónoma del Estado de Hidalgo,
Carretera Pachuca-Tulancingo Km. 4.5, C.P. 42184, Pachuca, Hgo.}

\author{Diego Portillo-S\'anchez}
\email{diego.portillo@cinvestav.mx}
\affiliation{Departamento de F\'isica, Centro de Investigaci\'on y de Estudios Avanzados del Instituto Polit\'ecnico Nacional
{\it Apartado Postal 14-740, 07000 Ciudad de M\'exico, M\'exico}}

\author{Genaro Toledo}%
 \email{toledo@fisica.unam.mx}
\affiliation{Instituto de F\'{\i}sica,  Universidad Nacional Aut\'onoma de M\'exico, Apartado Postal 20-364,  Ciudad de M\'exico 01000, M\'exico}%

\date{\today}% It is always \today, today,
             %  but any date may be explicitly specified

\begin{abstract}

We study the four-body heavy baryon decay, including two leptons in the final state, of the form $B_A \to B_B P \ell_\alpha \ell_\beta$, which can be either a lepton number conserving (LNC) or a lepton number violating (LNV) process (where $B$, $P$ and $\ell$ are baryons, pseudoscalar mesons and leptons, respectively), including all kinematic allowed lepton pair possibilities. We work beyond the simplified assumption of a single heavy Majorana neutrino mixing with the active sector, considering potential interference effects when including two nearly degenerate heavy  Majorana neutrinos. Particularly, we provide a first estimate of the decay channels involving different flavors for the external leptons (e.g., muon-tau), and elaborate on the interference pattern due to leptons exchange. We show that the results for the two heavy almost degenerate neutrinos can be recast into a single Majorana neutrino, and exhibit the features for both the LNC and LNV scenarios. 
We determine the potential exclusion region for the mass and heavy-light mixing parameters, of the neutrinos driving the decay, including finite size detector effects, and compare them with those from direct searches and $\tau$ and meson decays.
We also obtain the Branching ratio as a function of the heavy neutrino mass for the current upper limit of the heavy-light mixings.

\end{abstract}

\maketitle

\section{Introduction} \label{sec:intro}

Heavy neutral lepton states are a key ingredient in the understanding of light neutrino masses. The existence of such states would trigger new phenomena, such as lepton flavor violating (LFV) and LNV processes, which must be searched for to validate it. Perhaps, the simplest description of LNV processes (with the lepton number, $L$, changing in two units), includes only a single new neutral state, necessarily a neutrino of Majorana nature \cite{Atre:2009rg,Akhmedov:2014kxa,Abada:2017jjx}. On the other hand, LNC processes, can be driven by heavy neutrinos, which can be either Dirac or Majorana particles. This motivates the search not only for direct signals of these heavy neutral leptons \cite{Feng:2024zfe} but also for their properties, trying to distinguish between the two possible neutrino natures in various complementary processes (see, e.g., \cite{Marquez:2024tjl}). The approach with a single heavy neutral state provides a general description of the main features of LNC or LNV processes \cite{Gribanov:2001vv,Hernandez-Tome:2022ejd,Atre:2009rg, Helo:2010cw, Quintero:2011yh,Castro:2013jsn,Mejia-Guisao:2017nzx,Cvetic:2017vwl,Bar-Shalom:2006osy}. However, more realistic models (motivated to accommodate the mass generation mechanism \cite{Abada:2022wvh,CentellesChulia:2024uzv}) invoke two almost degenerate Majorana neutrinos, with masses in the GeV region. A recast of the single Majorana fermion results allows one to leverage previous calculations to determine the corresponding bounds for this case. Moreover, this extension incorporates relative phases that can greatly modify the results of a single neutrino, opening the door to explore also new phenomena, such as CP-violation \cite{Boyanovsky:2014una,Cvetic:2015naa,Cvetic:2023bwr,Das:2021kzi,Das:2021prm,Piazza:2022yrw,Abada:2022wvh, Najafi:2020dkp}.\\
This idea of having two almost degenerate neutrinos has been explored recently, in three-body \cite{Abada:2019bac, Zhang:2020hwj, Godbole:2020jqw, Cvetic:2023bwr} and four-body decays \cite{Cvetic:2015naa,Cvetic:2013eza,Das:2021prm} meanly for processes with same-flavor charged leptons produced at final state. 
On the other hand, the case of different flavors requires careful handling, associated with the leptons exchange. A prescription to deal with different leptons, in three-body decays, was presented in  \cite{Atre:2009rg, Abada:2017jjx}. Such procedure has the advantage of allowing a clean description of the individual channels interference, upon the leptons interchange. In a recent work, the integration method for four-body decays was presented \cite{Hernandez-Tome:2022ejd}, extending the case three-body decay and allowing us to study processes as the ones considered here.\\
Currently, the light charged leptons sector has the most restrictive bounds, for both the heavy-light mixing parameters and the region of the heavy neutrino mass. In contrast, for heavy-light mixing with tau flavor, the current bounds are less stringent than those in the light sector and arise, mainly, from the semileptonic tau decays \cite{Abada:2017jjx, Cvetic:2023bwr}. In this direction, the heavy baryon sector naturally allows the study of semileptonic processes, where the tau can be produced in combination with a light lepton. Thus, they may offer an opportunity to improve or complement the current constraints obtained from other sorts of processes.\\
In this work, we compute the four-body LNV decays of heavy baryons $B_A^{\{0,\mp\}} \to B_B^{\{\pm,0\}}P^{\pm} \ell_\alpha^\mp \ell_\beta^\mp$,
 and the LNC decays $B_A^{\{0,\mp\}} \to B_B^{\{\pm,\, 0\}}P^{\mp} \ell_\alpha^\mp \ell_\beta^\pm$,
 where $B$, $P$ and $\ell$ are baryons, pseudoscalar mesons and leptons, respectively. We follow the description of two almost degenerate heavy Majorana neutrinos as done by Ref.~\cite{Abada:2022wvh}, considering a set of form factors parameterizing the hadronic transition provided in the literature \cite{Ke:2019smy,Ke:2024aux,Zhao:2018zcb}. The systems we consider are generically associated to the following states $\Lambda_b \to (\Lambda_c,p) (K,\pi) \ell_\alpha \ell_\beta$, $\Sigma_b \to \Sigma_c (K,\pi) \ell_\alpha \ell_\beta$, 
$\Xi_b \to (\Sigma,\Xi_c,\Lambda) (K,\pi) \ell_\alpha \ell_\beta$ and
$\Omega_b \to (\Xi,\Omega) (K,\pi) \ell_\alpha \ell_\beta$, 
where $ \ell_\alpha, \ \ell_\beta = e, \mu, \tau$.\\
The remainder of the manuscript is structured as follows: section \ref{sec:setup} sets the notation and describes the formalism for the weak interaction of the heavy neutrinos \cite{Abada:2022wvh}. In section \ref{sec:amplitudes}, we outline the LNC and LNV processes, obtaining the corresponding amplitudes. We also describe the relation between the single-heavy neutrino and the two almost degenerate neutrinos, the so-called recast function. We include the detector effects in section \ref{sec:detector}. In section \ref{sec:results} we present the results and discussion. Finally, in section \ref{sec:conclusions} we draw our conclusions. 

\section{THE SET UP} \label{sec:setup}

To describe the processes in the leptonic part, we use the same approach as in Ref.~\cite{Abada:2022wvh}. Specifically, we work within the framework of simplified SM extensions that involve the addition of $N$ extra neutral Majorana fermions, without making any assumptions about the mechanism responsible for neutrino mass generation (i.e., treating neutrino masses and lepton mixings as independent). In such a case, the leptonic charged current is modified as follows:

\begin{eqnarray}
\mathcal{L}_{ \textrm{c.c}}&=&-\frac{g}{\sqrt{2}}  
U_{\alpha i}\,\bar{\ell}_\alpha\gamma_\mu P_L\nu_i W_\mu^- +\textrm{h.c.},
\label{chcurrent}
\end{eqnarray}
where $P_{L}=(1-\gamma_5)/2$ is the left-handed chirality projector, the subindex $i$ refers to the physical neutrino states (3 light plus $N$ heavy states), and the subindex $\alpha$ represents the flavor of the charged leptons. For the case $N=2$ (after the addition of two states with masses $m_{4,5}$), the unitary matrix $U$ has dimension $5\times 5$ encoding the flavour mixing ($U_{\alpha i}$) in charged currents. We employ the conventions presented in reference \cite{Abada:2022wvh} to denote the matrix elements for the heavy-light mixings by 
\begin{equation}
    U_{\alpha i}=e^{i\phi_{\alpha i}} |U_{\alpha i}|, \hspace{4mm} \alpha=e,\mu,\tau, \hspace{2mm} i=4,5\,
\end{equation} 
where $\phi_{\alpha i}$ is the phase of the associated mixing element.\\

\section{AMPLITUDES} \label{sec:amplitudes}

\begin{figure*}[]
\begin{center}
\begin{tabular}{c c c}
\includegraphics[scale=.5]{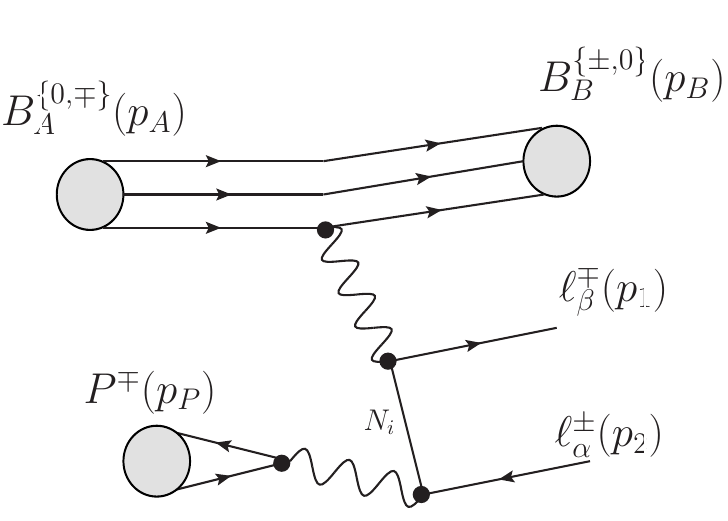}$\quad\quad\quad\quad\quad\quad\quad\quad\quad$& &\includegraphics[scale=.5]{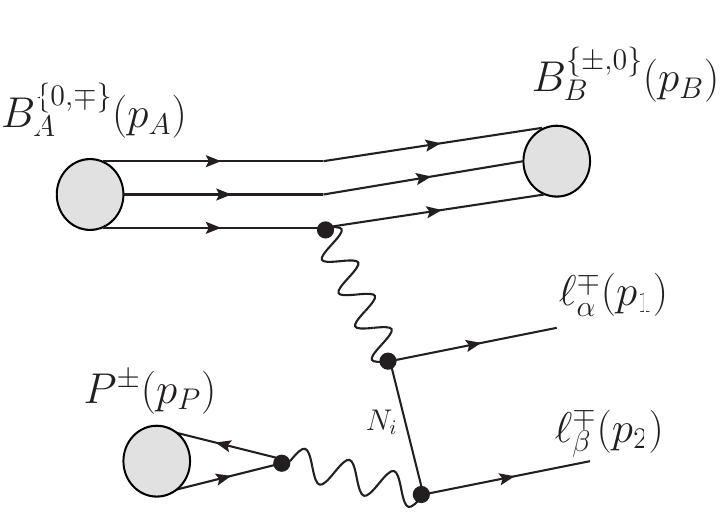}\\ (a) & & (b)
\end{tabular}
\caption{Feynman diagrams for the four-body semileptonic decays of heavy baryons mediated by a hypothetical resonant neutrino state.  The diagram (a) stands for the LNC case, namely, $B_A^{\{0,\mp\}} \to B_B^{\{\pm,0\}} P^\mp \ell_\alpha^\mp \ell_\beta^\pm$, whereas the diagram (b) represents the LNV scenario $B_A^{\{0,\mp\}} \to B_B^{\{\pm,0\}}P^{\pm} \ell_\alpha^\mp \ell_\beta^\mp$. In both cases,  $P$ is a pion or kaon state.}
\label{diagrams}
\end{center}
\end{figure*}

The dominant Feynman diagrams for the four-body semileptonic decays of heavy baryons, mediated by a hypothetical resonant neutrino state, are shown in Fig.~\ref{diagrams}. 
   The diagram \ref{diagrams}(a) stands for the LNC case, whereas the diagram \ref{diagrams}(b) represents the LNV scenario.
 We proceed to obtain the corresponding amplitudes.

\subsection{LNC \texorpdfstring{$B_A^{\{0,\mp\}} \to B_B^{\{\pm,\, 0\}}P^{\mp} \ell_\alpha^\mp \ell_\beta^\pm$}{BA-BbPll} DECAYS}

Let us start by considering the LNC case (diagram \ref{diagrams}(a)). Within these processes, we will focus on lepton flavor violating (LFV) decays, i.e., with $\ell_\alpha\neq\ell_\beta$, to isolate the sterile neutrino contributions (decays with the same-flavor lepton final states also receive a SM contribution). We will obtain the amplitude for the specific decay mode $B_A^{\{0,-\}}(p_A) \to B_B^{\{+,\, 0\}}(p_B)\ P^{-} (p_P) \ell_\beta^-(p_1) \ell_\alpha^+(p_2)$, where in parentheses are the corresponding momenta. An analogous procedure can be followed to obtain the charge conjugate modes. Using the charged-current Lagrangian in Eq. (\ref{chcurrent}), the amplitude can be written as follows:
\begin{align} \label{amp-com-1}
\mathcal{M}_{\text{LNC}}&=G\,p_{P}^\nu \sum_{i=4,5} \,U_{{\beta}i}\,U^*_{{\alpha}i}\, \ell^{\rm{LNC}}_{\mu\nu}\, P_{1i} \, H^\mu(p_B, p_A)\,,
\end{align}
where we have defined $G\equiv G_F^2V_{AB}V_{P} f_{P}$, with $V_{P}$ and $V_{AB}$ the mixing quark elements of the CKM matrix involved in the hadronic part of the process, and $f_P$ the decay constant of the meson state (for $P=\pi, K$ they correspond to $V_P=V_{ud}, V_{us}$ and  $f_P=90, 120$ MeV, respectively). The leptonic part is given by
\eq{
\ell_{\mu\nu}^{\rm{LNC}}&\equiv \bar{u}(p_1) \gamma_\mu \slashed{a}_1  \gamma_\nu (1-\gamma_5) v(p_2),
}
where $a_{1}\equiv p_A-p_B-p_{1}$ is the momentum carried out by either of the heavy neutrinos, and we have defined 
\eq{
P_{1i}\equiv \frac{1}{a_1^2-m_{i}^2+im_{i}\Gamma_{i}}.
\label{eq:P1i}
}

The hadronic part is given by the meson production coming from the $W$, parameterized by $if_P p_P^\nu$, already included in Eq.  (\ref{amp-com-1}) and the baryons transition matrix element 
\eq{
H^\mu(p_B,p_A)&\equiv \langle B_B(p_B) \vert J^\mu  \vert B_A(p_A)\rangle.\label{Hmu}
}
The most general form is parameterized by six form factors, as follows 
\eq{
&\langle B_B(p_B)\vert  J_\mu \vert B_A(p_A) \rangle = \nonumber\\
&\bar{u}(p_B)\bigg[ f_1(q^2)\gamma_\mu +i f_2(q^2) \frac{\sigma_{\mu\nu}q^\nu}{M_A}+\frac{q_\mu f_3(q^2)}{M_A}\label{ff}\\
&+ g_1(q^2)\gamma_\mu\gamma_5 +i g_2(q^2) \frac{\sigma_{\mu\nu}q^\nu \gamma_5}{M_A}+\frac{q_\mu g_3(q^2) \gamma_5}{M_A}\bigg] u(p_A)\nonumber,
}
with $q^2=(p_A-p_B)^2$ the squared momentum transferred in the baryonic transition, whereas $u(p_A)$ ($\bar{u}(p_B)$) and $M_A$  ($M_B$) are the spinor and mass of the initial (final) baryon.
The form factors have been computed
considering specific approaches to quark models and/or lattice calculations \cite{ Detmold:2015aaa,Neishabouri:2024gbc,Zhao:2018zcb, Ke:2019smy, Ke:2024aux, Miao:2022bga}.
To avoid model-dependent effects coming from different estimations of the form factors, we use those obtained in the light-front model \cite{Zhao:2018zcb, Ke:2019smy, Ke:2024aux}, for all baryonic transitions considered in this work. In the approach used in those references the $f_3$ and $g_3$ form factors cannot be extracted; however, they vanish in the heavy quark limit\cite{Ke:2007tg}. Therefore, we only consider the $f_1, f_2, g_1$ and $g_2$ form factors, as given in Appendix \ref{ap:ff}. The contribution of $f_2$ and $g_2$ to the decay width is subdominant, compared to the vector and axial couplings $f_1$ and $g_1$. However, we incorporate them in the numerical evaluation for a complete description of the effective baryon weak transition. Different parameterizations of the form factors, as those reported in Ref.\cite{Detmold:2015aaa}, have been used to determine their effect on the observables. No significant changes were observed in the final results.\\
The squared amplitude  for the LNC case is given by
\begin{align}
|\mathcal{M}_{\text{LNC}}|^2&=
\Big(
|U_{\alpha 4}|^2 |U_{\beta 4}|^2 |P_{14}P_{14}^*
+
|U_{\alpha 5}|^2 |U_{\beta 5}|^2 |P_{15}P_{15}^*
\nonumber\\
&+ U_{\alpha 4}^* U_{\beta 4} 
U_{\alpha 5} U_{\beta 5}^*
P_{14}P_{15}^*
+
U_{\alpha 4} U_{\beta 4}^* 
U_{\alpha 5}^* U_{\beta 5}
P_{14}^*P_{15}
\Big)\nonumber\\
&G^2 \ell_{\mu\nu}^{\rm LNC} p_P^\nu 
 \ell_{\tau\theta}^{\dagger {\rm LNC}} p_P^\theta H^\mu {H^{\tau}}^\dagger,
\label{ASLNC} 
\end{align} 
where the interference terms can be stated in terms of the relative phases as

\begin{align}
   & U_{\alpha 4}^* U_{\beta 4} 
U_{\alpha 5} U_{\beta 5}^*
P_{14}P_{15}^*
+
U_{\alpha 4} U_{\beta 4}^* 
U_{\alpha 5}^* U_{\beta 5}
P_{14}^*P_{15}=\nonumber\\
&\hspace{7mm}|U_{\alpha 4}| |U_{\beta 4}| 
|U_{\alpha 5}| |U_{\beta 5}|
\Big(
e^{i(\psi_{\alpha}- \psi_{\beta})}
P_{14}P_{15}^*\nonumber\\
&\hspace{3.6cm}+
e^{-i(\psi_{\alpha}- \psi_{\beta})}
P_{14}^*P_{15}
\Big)
\end{align}
where
$\psi_{\alpha}\equiv \phi_{\alpha 5}-\phi_{\alpha 4} $. 
This squared amplitude can be recast into a single Majorana neutrino process, provided the masses satisfy $m_4\simeq m_5\equiv m_N$ and $\Delta m_N\equiv m_5-m_4 \gtrsim 0$, the decay widths are $\Gamma_4 \simeq \Gamma_5 \equiv \Gamma_N$, the mixing parameters are thus also required to fulfill $|U_{\alpha4}||U_{\beta4}|=|U_{\alpha5}||U_{\beta5}|=|U_{\alpha N}||U_{\beta N}|$
and the relation between the $P_{1i}P_{1j}^*$, is as shown in Appendix \ref{ap:recast}. Thus, the average square amplitude becomes:

\begin{align}
\overline{|\mathcal{M}_{\text{LNC}}|^2}&=
G^2 \ell_{\mu\nu}^\text{LNC} p_P^\nu 
 \ell_{\tau\theta}^{\dagger {\rm LNC}} p_P^\theta H^\mu {H^{\tau}}^\dagger \,|U_{\alpha N}|^2 |U_{\beta N}|^2 \nonumber\\
&\frac{\pi}{2m_N \Gamma_N}\delta(a_1^2-m_N^2) R[y,\psi_-],\label{eq:amplitudeLNC}
\end{align}
where $y=\Delta m_N/\Gamma_N$ and $R[y,\psi]$ is the recast function
\begin{align}  
R[y,\psi_\mp]\equiv
2\Big\{1+ \kappa(y)\big(
{\rm cos}(\psi_\alpha \mp \psi_\beta)-y\,{\rm sin}(\psi_\alpha\mp \psi_\beta)
\big)\Big\},
\label{eq:LNCrecast}
\end{align}
with $\kappa(y)=1/(1+y^2)$. 
The $R[y,\psi_+]$ function corresponds to the LNV case, as we show later.
From Eq.(\ref{eq:amplitudeLNC}) we see that the effect of the extended scenario with two neutrinos (in the quasi-degenerate limit) on the single neutrino scenario is included in the recast factor $R$. Therefore, it is enough to compute the square amplitudes, for the case of a single resonant neutrino and multiply it by the recast function to get the full result, in the two almost degenerate case. However, the presence of this $R$ factor,  can strongly modify the magnitude of the amplitude. Therefore, it will have a direct impact on the constraints for the branching fraction and heavy-light mixing, as we discuss later.
For the LNC charge conjugate process, the argument of the $sine$ and $cosine$ functions changes from $\psi_\alpha - \psi_\beta \to -(\psi_\alpha - \psi_\beta)$, associated to the modification of the heavy-light mixing parameters involved. Then, the relative sign on the $sine$ function with respect to the $cosine$ one, in Eq. (\ref{eq:LNCrecast}), changes from negative to positive.

\begin{figure}[t]
\centering
\subfigure{\includegraphics[width=8cm]{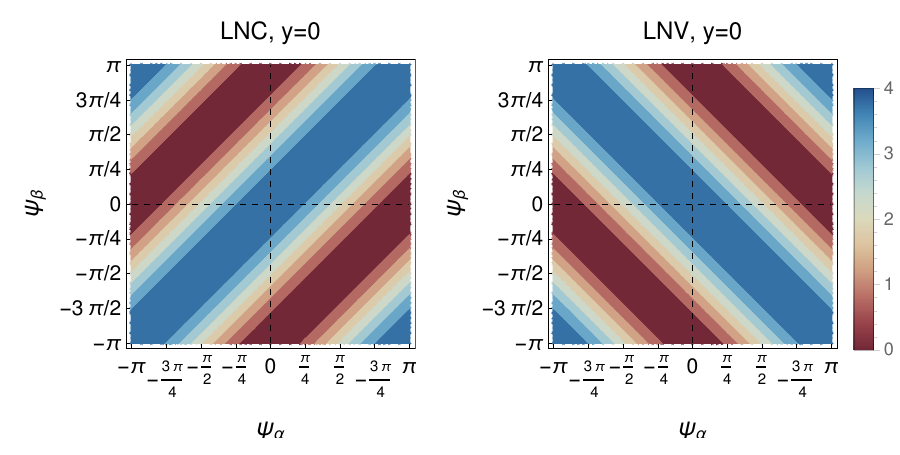}}
\subfigure{\includegraphics[width=8cm]{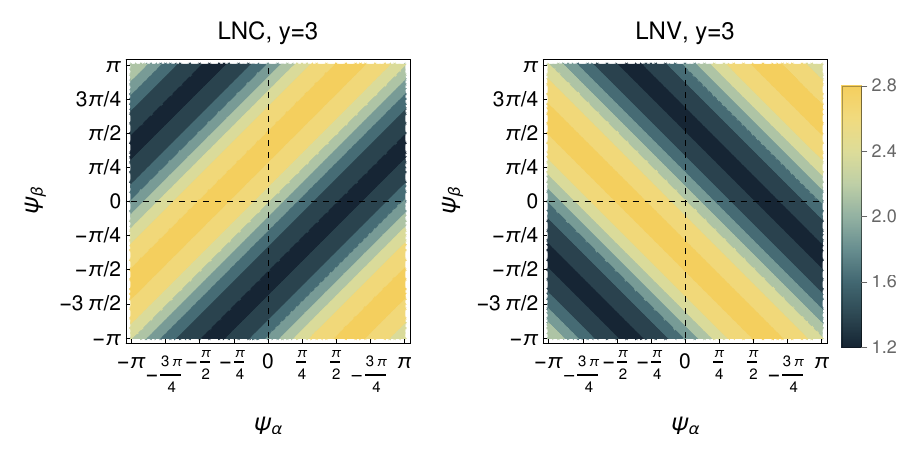}}
\subfigure{\includegraphics[width=8cm]{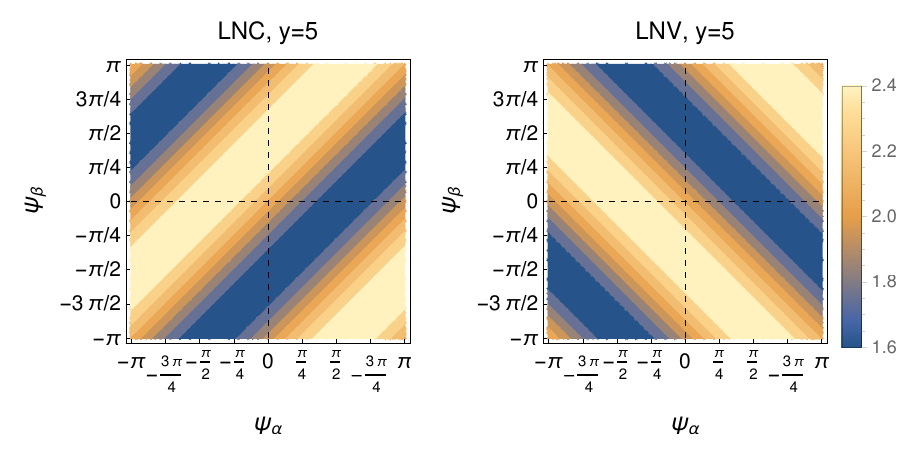}}
\caption{Recast function behavior,  Eq. (\ref{eq:LNCrecast}), for LNC and LNV processes as a function of the $\psi_\alpha$ and $\psi_\beta$ phases and using different values of $y=\Delta m_N/\Gamma_N$.}
\label{fig:recast}
\end{figure}

To better understand the role of the recast factor $R$, in Fig. \ref{fig:recast} we plot it as a function of the phases $\psi_\alpha$ and $\psi_\beta$, for three values of $y$ (rows).
Notice that the range in which the recast function varies depends on the value of the $y$ parameter. For $y=0$ the range is maximum, $[0,4]$, and decreases to $2$ as $y$ increases. In fact, for large values of $y$, the interference effect becomes negligible, as pointed out in Ref. \cite{Abada:2019bac}. On the other hand, we also compare the behavior of the recast function for LNC (left column) and LNV (right column) processes. For $y=0$ there are scenarios where the LNV process has a maximum and the LNC one vanishes, for example, when $\psi_\beta=-\psi_\alpha=\pi/2$. The opposite scenario, i.e., the LNV process vanishes while the LNC one is maximum, occurs for $\psi_\beta=\psi_\alpha=\pi/2$. This last case corresponds to the full degeneracy limit, where both Majorana fields form a Dirac singlet, recovering Lepton Number as an exact symmetry.    
\\

\subsection{LNV \texorpdfstring{$B_A^{\{0,\mp\}} \to B_B^{\{\pm,0\}}P^{\pm} \ell_\alpha^\mp \ell_\beta^\mp$}{BA-BbPll} DECAYS} 

Regarding the LNV case, depicted in Fig. \ref{diagrams}(b), we must include an additional diagram associated with the interchange of the final charged leptons, regardless of their flavor. We will obtain the amplitude  for the specific decay mode $B_A^{\{0,+\}}(p_A) \to B_B^{\{-,\, 0\}}(p_B)\ P^{-} (p_P) \ell_\alpha^+(p_1) \ell_\beta^+(p_2)$. The amplitude is given by
\begin{align} \label{amp-com-2}
    \mathcal{M}_{\text{LNV}}&= G\,p_{P}^\nu \sum_{i} \, U^*_{\alpha i}\,U^*_{\beta i} \,m_{i}\, H^\mu(p_B, p_A)\, \nonumber\\
    &\Big(\ell^{\rm{LNV}}_{\mu\nu}(p_1,p_2) P_{1i}+ \ell^{\rm{LNV}}_{\nu\mu}(p_1,p_2) P_{2i}\Big)\,,
\end{align}
where the leptonic part in this case is written as 
\eq{
\ell_{\mu\nu}^\textrm{LNV}(p_1,p_2)\equiv 
\bar{u}(p_1) \gamma_\mu \gamma_\nu (1+\gamma_5) v(p_2),
}
and $P_{2i}$ is similar to Eq. (\ref{eq:P1i}) with $a_1 \to a_2$, $a_2=p_A-p_B-p_2$.
The Fermi quantum statistics properties for identical leptons in the final state yield the anti-symmetrization of the second amplitude in Eq.(\ref{amp-com-2}); meanwhile, in the case of different flavors the minus sign appears after the anti-commutation relations of the fermion fields, i. e. an odd permutation in the order of the spinors at the amplitude level, as is well explained in Ref.\cite{Denner:1992vza}. Then, in Eq. (\ref{amp-com-2}) we have a plus sign between the leptonic parts, since $\ell_{\mu\nu}^\text{LNV}(p_2,p_1)=-\ell_{\nu\mu}^\text{LNV}(p_1,p_2)$, obtained by applying charge-conjugation relations.

The hadronic part is given by Eq. (\ref{ff}), presented in the LNC case, with the same considerations. The square amplitude is given by

\begin{widetext}
\begin{align}
|\mathcal{M}_{\text{LNV}}|^2&= G^2 p_P^\nu p_P^\theta H^\mu {H^{\tau}}^\dagger \nonumber\\
&\times \Bigg\{ \Big[ \sum_{i=4,5}
  m_i^2
|U_{\alpha i}|^2 |U_{\beta i}|^2 P_{1i}P_{1i}^*  
+2 m_4 m_5 \,{\rm Re}(U_{\alpha 4}^* U_{\beta 4}^* 
U_{\alpha 5} U_{\beta 5}P_{14}P_{15}^*) \Big]
 \ell_{\mu\nu}^{\rm LNV}  
 \ell_{\tau\theta}^{\dagger {\rm LNV}} \nonumber\\
& 
+ \Big[ \sum_{i=4,5}
 m_i^2
|U_{\alpha i}|^2 |U_{\beta i}|^2 P_{2i}P_{2i}^* +2 m_4 m_5 \,{\rm Re}(U_{\alpha 4}^* U_{\beta 4}^* 
U_{\alpha 5} U_{\beta 5}P_{24}P_{25}^*)
\Big] \ell_{\nu\mu}^{\rm LNV}  
 \ell_{\theta\tau}^{\dagger {\rm LNV}} \nonumber\\
&+\Big[\sum_{i=4,5} m_i^2 |U_{\alpha i}|^2 |U_{\beta i}|^2 P_{1i} P^*_{2i}
+\sum_{(i\neq j)=4,5} m_i m_j U_{\alpha i}^* U_{\beta i}^* 
U_{\alpha j} U_{\beta j}\,P_{1i}P_{2j}^* \Big] \ell_{\mu\nu}^{\rm LNV}  
 \ell_{\theta\tau}^{\dagger {\rm LNV}} \nonumber\\
& +\Big[\sum_{i=4,5}
 m_i^2 
|U_{\alpha i}|^2 |U_{\beta i}|^2  \,P_{2i}P_{1i}^*
+ \, \sum_{(i\neq j)=4,5}m_i m_j\,U_{\alpha i}^* U_{\beta i}^* 
U_{\alpha j} U_{\beta j} \,P_{2i}P_{1j}^* \Big]
\ell_{\nu\mu}^{\rm LNV}  
 \ell_{\tau\theta}^{\dagger {\rm LNV}}\Bigg\}
\label{ASLNV} 
\end{align} 
\end{widetext}

Some observations are in order for the four-body decay:
to set the squared amplitude in the recast form, we make use of the general results as presented in Appendix B.
From Eq. (\ref{ASLNV}), we can notice that the individual channels contribution (first and second row) are similar to the LNC squared amplitude. 
\begin{align} \label{eq:amplitudeLNV}
\to\, & G^2 p_P^\nu p_P^\theta H^\mu {H^{\tau}}^\dagger    
 \frac{\pi m_N}{2\Gamma_N}
|U_{\alpha N}|^2 |U_{\beta N}|^2   R[y,\psi_+] \nonumber\\
 &\Bigg\{
 \ell_{\mu\nu}^{\rm LNV}  
 \ell_{\tau\theta}^{\dagger {\rm LNV}}\delta(a_1^2-m_N^2)
+ \ell_{\nu\mu}^{\rm LNV}  
 \ell_{\theta\tau}^{\dagger {\rm LNV}}\delta(a_2^2-m_N^2) \Bigg\}\nonumber\\
 \end{align} 

Therefore, the recast function takes the same form as Eq. (\ref{eq:LNCrecast}) with $\psi_\alpha-\psi_\beta \to \psi_\alpha+\psi_\beta$. This result is valid regardless of the flavor of the leptons. The remaining part, the channels interference must be computed using the relations as given in Appendix \ref{ap:recast}. If the leptons are of different flavor, they have different kinematical regions for the neutrino mass value; if such individual kinematical regions do not overlap, the interference is null. The only case under consideration where the kinematical regions overlap, is in the $\Lambda_b \to p (\pi, K) \mu^- \tau^-$ decay. We computed the interference contribution and find it very small compared to the individual channels.\\

\section{Detector length considerations} \label{sec:detector}

An important effect to be taken into account in the calculation is the probability of the on-shell neutrino decaying inside the detector; otherwise, the signal event will correspond only to the first sub-process without any meaningful information about the resonant state. As a consequence, the detector consideration will directly affect the constraint on the branching fraction (and consequently to the heavy-light mixings), mainly for neutrinos with masses smaller than $\sim 500$ MeV (due to the large values of their lifetimes), as we will discuss later. 

From the experimental point of view, the neutrino lifetime range would correspond with a decay distance from the collision point, within the detector’s sensitivity range~\cite{CMS:2014pgm}.
Considering the LHC experimental setup, as a reference for a detector, a constant lifetime of the order $1-1000$ ps has been used~\cite{Mejia-Guisao:2017nzx}.
However, we adopt a different strategy, that allows for a more accurate evaluation of $\Gamma_N$, considering it as a function of the heavy-light mixings and the mass of the heavy state\footnote{ We take the sum over exclusive decay channels of the heavy state, i. e. $\Gamma_N=\sum_i \Gamma^{p.w}_i \, \theta(m_N - \sum_j m_j)$, where $\theta$ is the Heaviside function and $m_j$ stand for the masses of all the final states
particles involved in the partial width $\Gamma_i ^{p.w}$.}
Therefore, following Ref.~\cite{Abada:2017jjx}, for the scenario with just one heavy neutrino state, the detector effect is encompassed in the following probability weight:
\eq{
P_{\nu}=1- {\rm Exp}\left(-L_{\textrm{det}}\, \Gamma_N \frac{m_N}{|p_N|}\right)\label{eq:decayingprobability},
}
where $L_{\textrm{det}}$ is the detector length, $\Gamma_N$ the total decay width of the heavy neutrino and $p_N$ its three-momentum in the laboratory frame of the decaying baryon~\footnote{For simplicity, we take the scenario where the initial baryon is produced at rest, however, this can be generalized to the case where the decaying particle is boosted, as done in Ref.\cite{Cvetic:2023bwr}.}.
Therefore, the differential decay rate can be written as:
\eq{
d\Gamma=P_{\nu}|\overline{\mathcal{M}|^2}\,dPS,
} being $dPS$ the corresponding Phase Space integral according to the transition. In a three-body LNV decay (such as $M_1^-\to \ell_1^- N \to M_2^+\ell_1^-\ell_2^-$), after applying the narrow width approximation, the probability $P_{\nu}$ becomes a constant given that the first sub-process ($M_1^-\to\ell_1^-N$) corresponds to a two-body decay with a monochromatic spectrum (where the momentum $|p_N|=\lambda^{1/2}(m_{M_1}^2,m_N^2,m_{\ell_1}^2)/(2 m_{M_1})$ \footnote{With $\lambda(a,b,c)\equiv a^2+b^2+c^2-2(a b+a c+b c)$.} is fixed by energy-momentum conservation). Therefore the decay rate is computed straightforwardly:
 \eq{
\Gamma=P_{\nu}\int|\overline{\mathcal{M}|^2} dPS.
}
On the contrary, for a four-body decay (for instance, $B_A^{0}\to B_B^{\pm}\,\ell_1^{\mp}N\to B_B^{\pm}\ell_1^{\mp}\ell_2^{\mp} P^{\pm}$), the momentum $|p_{N}|$ is a function of the invariant mass of the $B_B-\ell_1$ system, and will depend on the phase space variables even after the narrow-width approximation. In this sense, it is convenient to write the total decay width as:
\eq{
\Gamma=&\Gamma_0-\textrm{Exp}(-2L_{\textrm{det}}\Gamma_N m_N)\nonumber\\
&\times \int \textrm{Exp}\left(\frac{m_{A}}{\lambda^{1/2}(m_A^2,s_{Bi},m_N^2)}\right) f_{PS_i} dPS_i,\label{eq:detectoreffect}
} where the sum over $i={1,2}$ is implicit. $s_{Bi}\equiv(p_B+p_i)^2$ and $\Gamma_{0}$ is the decay width of an infinite detector length, that is, $P_{\nu}=1$. The $f_{PS_i}$ functions are defined as follows (see Eq. (2.9) in Ref.\cite{Hernandez-Tome:2022ejd}):
\eq{
f_{PS_i}=\frac{\overline{|\mathcal{M}_i|^2}}{\overline{|\mathcal{M}_1|^2}+\overline{|\mathcal{M}_2|^2}}|\overline{\mathcal{M}|^2},
} 
being $\mathcal{M}_{1}$ the amplitude of the diagram in Fig. \ref{diagrams}(b), and $\mathcal{M}_2$ the one for the diagram with exchanged leptons. The definition of the phase space $PS_i$ is given in Appendix A of Ref.~\cite{Hernandez-Tome:2022ejd}, where $dPS_1=dPS_2(p_1\leftrightarrow p_2)$. On the other hand, for the LNC processes under consideration, the absence of the diagram with exchanged leptons in the final state yields to i=1 and, therefore, $f_{PS_1}=|\overline{\mathcal{M}|^2}$.\\

It is important to note that the above explanation is valid only in the case of a single heavy neutrino state (otherwise, we have to sum over each resonant neutrino state). However, the generalization to the quasi-degenerate limit of two Majorana neutrinos is straightforward, following the expressions in Eqs.~(\ref{eq:amplitudeLNC}) and (\ref{eq:amplitudeLNV}) and the proper incorporation of the recast factor.

\section{Results and Discussion} \label{sec:results}

To fully compute the decay processes, we consider the form factors for each case as given in Appendix \ref{ap:ff}. The neutrino decay width is taken as dependent on the heavy neutrinos mass, the heavy-light mixing parameters and the corresponding open channels. We follow Ref.~\cite{Atre:2009rg, Abada:2017jjx} to compute the total decay width as the sum over exclusive channels. 
In Table \ref{tab:processes}, we list the generic processes considered in this work, for either the $K$  or the $\pi$ meson in the final state  and the LNV or LNC lepton pair (where the transitions with two electrons in the final state are omitted, due to the strong constrains on the heavy-light mixing parameter), and the CKM element, $V_{AB}$, involved for each one.\\

\begin{table}[htb]
    \centering
    \begin{tabular}{lcc}
    \hline
    \hline
       Process &  $V_{AB}$ \\
    \hline 
    $\Lambda_b \to \Lambda_c (K,\pi) \ell_\alpha \ell_\beta$ & $V_{cb}$ \\
    $\Lambda_b \to p (K,\pi) \ell_\alpha \ell_\beta$   & $V_{ub}$ \\
        $\Sigma_b \to \Sigma_c (K,\pi) \ell_\alpha \ell_\beta$ & $V_{cb}$ \\
    $\Xi_b \to \Xi_c (K,\pi) \ell_\alpha \ell_\beta$   & $V_{cb}$\\
    $\Xi_b \to \Sigma (K,\pi) \ell_\alpha \ell_\beta$   & $V_{ub}$ \\
        $\Xi_b \to \Lambda (K,\pi) \ell_\alpha \ell_\beta$   & $V_{ub}$ \\
     $\Omega_b \to \Xi (K,\pi) \ell_\alpha \ell_\beta$  & $V_{ub}$\\
     $\Omega_b \to \Omega_c (K,\pi) \ell_\alpha \ell_\beta$ & $V_{cb}$\\
    \hline
    \hline
    \end{tabular}
    \caption{Particular processes considered in this work. For the leptonic pair we include $\ell_\alpha \ell_\beta$= $\mu e$, $\mu\mu$, $\tau e$ and $\tau \mu$.}
    \label{tab:processes}
\end{table}

\begin{figure*}[htp]
    \centering
    \begin{tabular}{cc}
         \includegraphics[scale=.59]{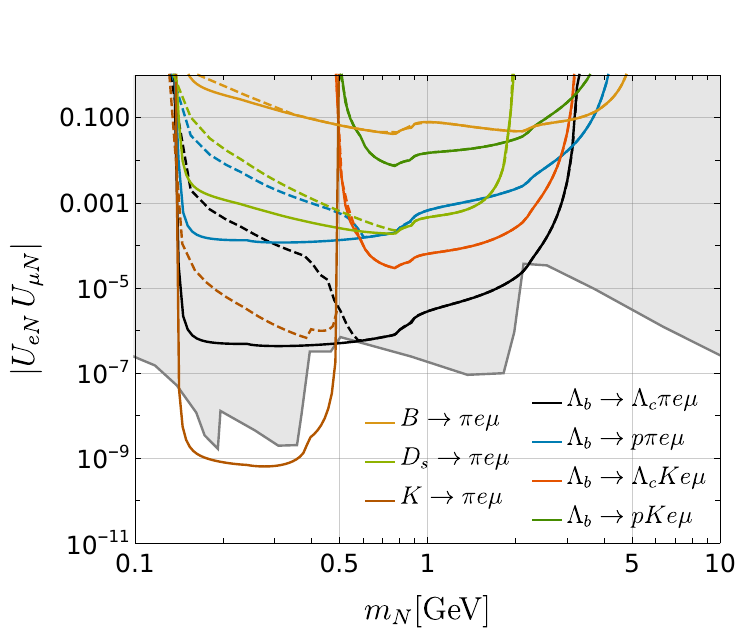} & \includegraphics[scale=.59]{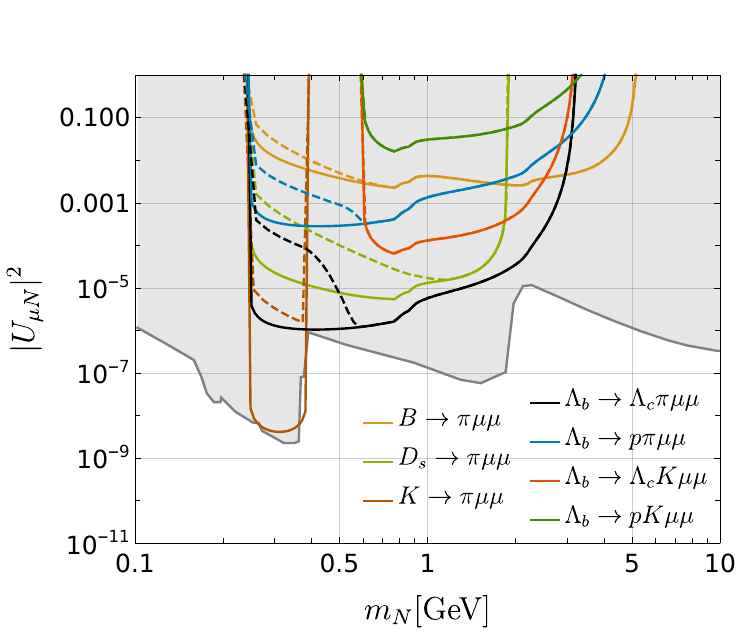}  \\
        \includegraphics[scale=.59]{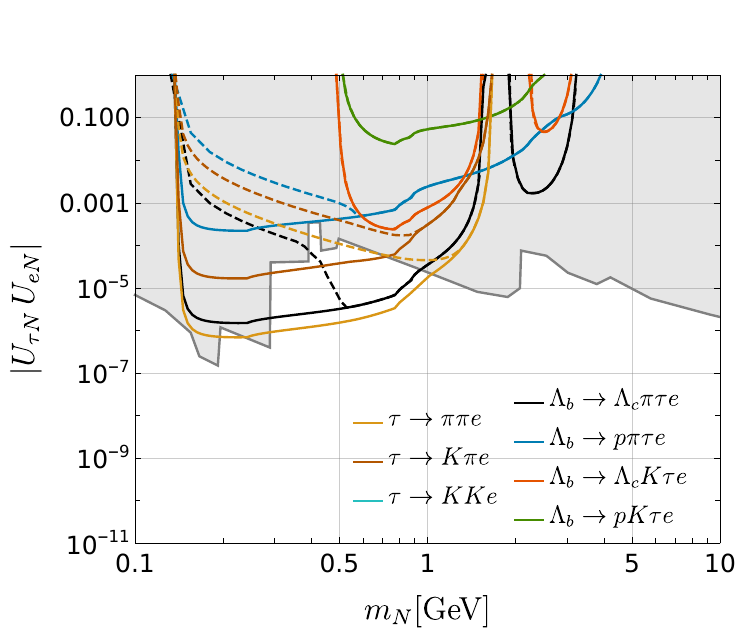}&\includegraphics[scale=.59]{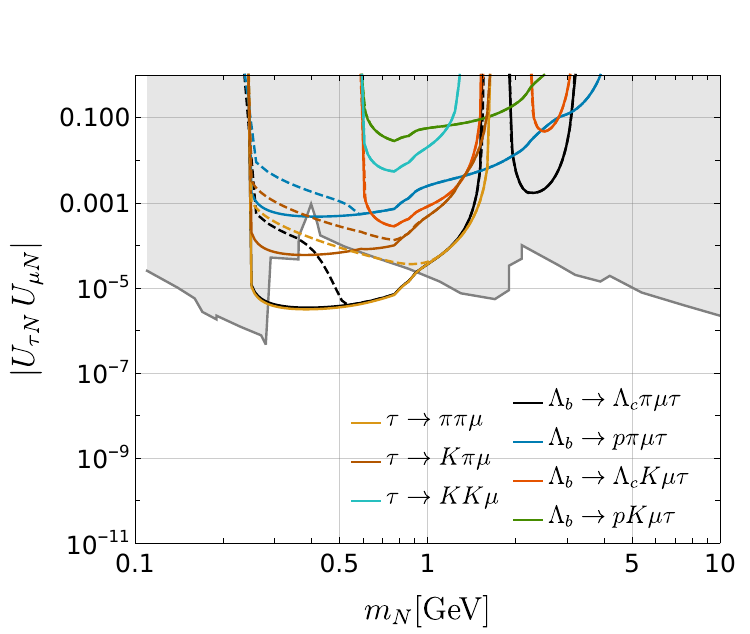}\\
    \end{tabular}
    \caption{Exclusion region of the heavy-light mixing, as a function of the heavy neutrino mass, for the LNV four-body $\Lambda_b$ decays. Considering the effect of a finite detector length ($L_{\rm det}=10$ m), yields the dashed lines. The exclusion regions for LNV $\tau^{\pm}\to M_A^\pm M_B^{\pm} \ell^\mp$ and meson $M_A^\pm \to M_B^{\mp} \ell_\alpha^\pm\ell_\beta^\pm$ decays, taken from Ref.\cite{Abada:2017jjx}, are also exhibited.  Shaded regions correspond to the limits from direct searches at colliders (obtained under different assumptions, see text for details) taken from \cite{Abada:2017jjx,Abada:2022wvh}.}
    \label{fig:LNVmixing}
\end{figure*}

For the sake of clarity, in the following we present the results for the LNV case with a single resonant neutrino. At this stage, the same results are obtained for the LNC case. Thus, the effect of having introduced two quasi-degenerate neutrinos is driven by the recast function, as discussed before. This is strictly true in the quasi-degenerate limit and narrow with approximation.\\
In order to simplify the numerical evaluation, we take the universal coupling assumption, i.e., we consider $|U_{eN}|=|U_{\mu N}|=|U_{\tau N}|$. In Fig. \ref{fig:LNVmixing}, we exhibit the exclusion region of the heavy-light mixings as function of the heavy neutrino mass for the LNV four-body $\Lambda_b$ decays under study, shown with solid lines. These regions are obtained under the assumption of an upper limit for the branching fraction of the four-body $\Lambda_b$ decays to be $\textrm{BR}\le\mathcal{O}(10^{-8})$, motivated by the expected sensitivity to the process $\Lambda_b\to\Lambda_c (p)\pi \mu\mu$ at CMS and LHCb experiments, analyzed in Ref.\cite{Mejia-Guisao:2017nzx}. Moreover, we incorporated the effect of a finite detector length ($L_{\rm det}=10$ m), according to Eq.(\ref{eq:detectoreffect}), represented with dashed lines in the same Fig. \ref{fig:LNVmixing}. There, we notice a clear underestimation of the exclusion regions for neutrinos with masses below $\sim 500$ MeV, compared with the case where no finite detector effects are included. 

For comparison purposes, we also plot the corresponding curves for LNV $\tau^{\pm}\to M_A^\pm M_B^{\pm} \ell^\mp$ as well as analogue exclusion regions coming from LNV meson $M_A^\pm \to M_B^{\mp} \ell_\alpha^\pm\ell_\beta^\pm$ decays, taken from Ref.\cite{Abada:2017jjx}.
Shaded regions correspond to the limits from direct searches at colliders, taken from \cite{Abada:2017jjx,Abada:2022wvh}, considering an extensive experimental information (see references in \cite{Abada:2017jjx}). We observe that the processes here considered may be as competitive as those from $\tau$ and meson decays.
It is important to note that, for direct searches, the limits on a given $U_{\ell N}$ mixing element were obtained 
by setting the other mixings with the heavy state to zero; for instance, the constraints on $|U_{\mu N}|^2$ are inferred assuming $U_{e N}=U_{\tau N}=0$. On the other hand, we work under the universal coupling assumption, as previously mentioned, meaning that the comparison between our results and the bounds from direct searches, in Fig. \ref{fig:LNVmixing}, must be taken with care, since they are not obtained under identical assumptions. Considering the same scenario as in direct searches, our bounds become in general more restrictive (by less than one order of magnitude), making them more competitive.\\
We also observe that, for a given process, the decay mode including the pion in the final state gives more restrictive bounds than the case including the Kaon, which is expected due to the values of the CKM elements and phase space suppression.

\begin{figure*}[htp]
    \centering
    \begin{tabular}{cc}
         \includegraphics[scale=.41]{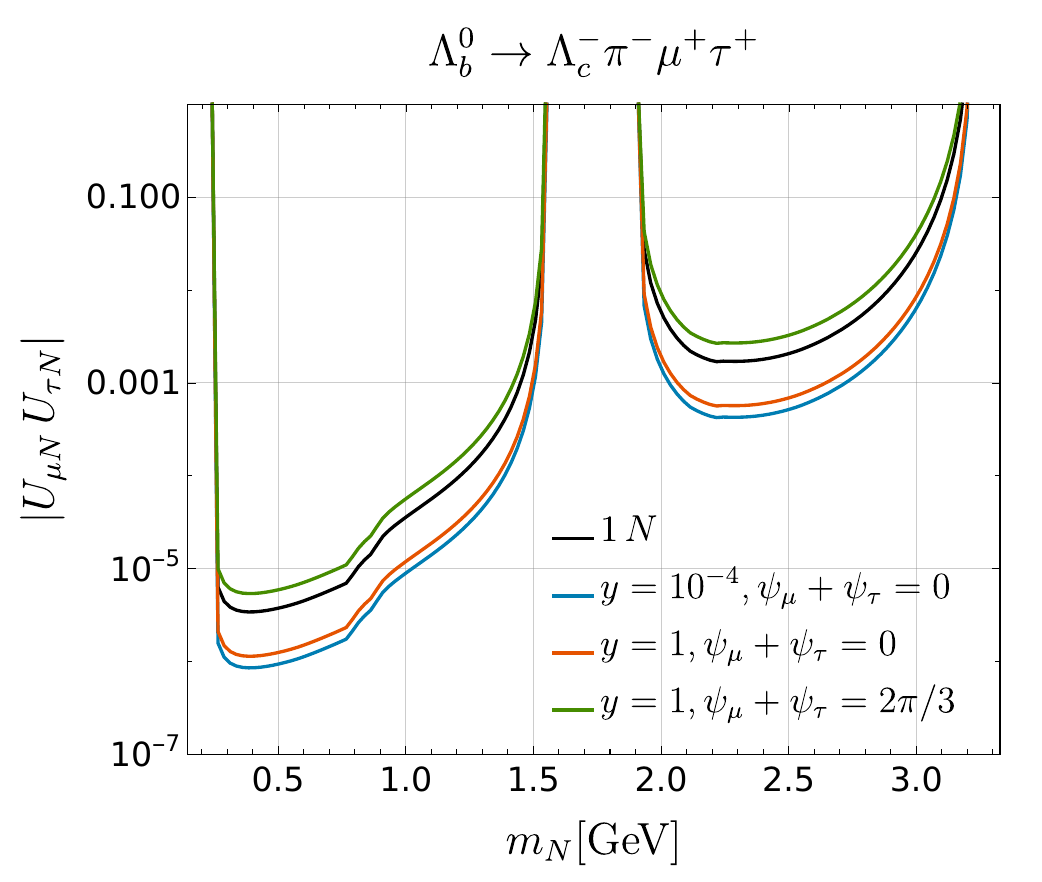}&
         \includegraphics[scale=.41]{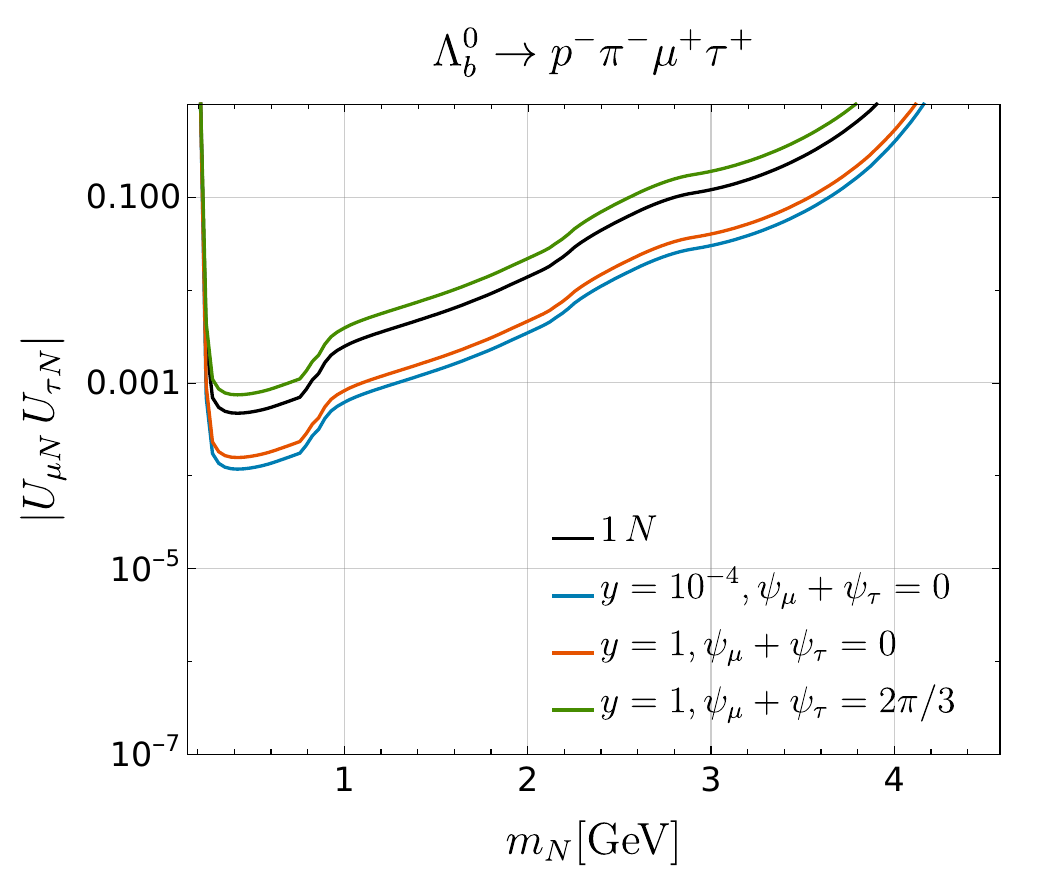} 
    \end{tabular}
    \caption{Impact on the upper limits for the heavy-light mixings for the $\Lambda_b^0\to\Lambda_c^-(p^-)\pi^-\tau^+\mu^+$ processes, after extending to the scenario with two quasi-degenerate Majorana neutrinos. For a set of values of $y$ and $\psi_{\mu}+\psi_{\tau}$. The black line ($1N$) corresponds to the case with only one heavy neutrino state, as reference.}
   \label{fig:LNVrecast}
\end{figure*}

We have explored the modifications on the upper limits, after extending to the scenario with two quasi-degenerate Majorana neutrinos. This is modulated by the recast function defined in Eq.~(\ref{eq:LNCrecast}). In Fig.~\ref{fig:LNVrecast}, we illustrate such an effect in the heavy-light mixing couplings for the $\Lambda_b^0\to\Lambda_c^+(p^-)\pi^-\tau^+\mu^+$ LNV processes, considering a set of values for the degeneracy parameter $y$ and the phase $\psi=\psi_\alpha+\psi_\beta$.  In order to compare the results in the extended scenario, we take the case with only one heavy neutrino state as a reference (denoted by $1N$ in the same figure). The behavior, for the different values of $y$ and the phase, can be understood using the recast function displayed in Fig.~\ref{fig:recast}. That is, the phase and $y$ can combine to make the impact on the upper limits either larger or less restrictive.
For instance, they can induce an extra suppression in the branching fraction of the process, making the exclusion regions for the heavy-light mixing smaller (which is the case for the green line in Fig.~\ref{fig:LNVrecast}).

\begin{figure*}[htp]
    \centering
    \begin{tabular}{c}
        \includegraphics[scale=.5]{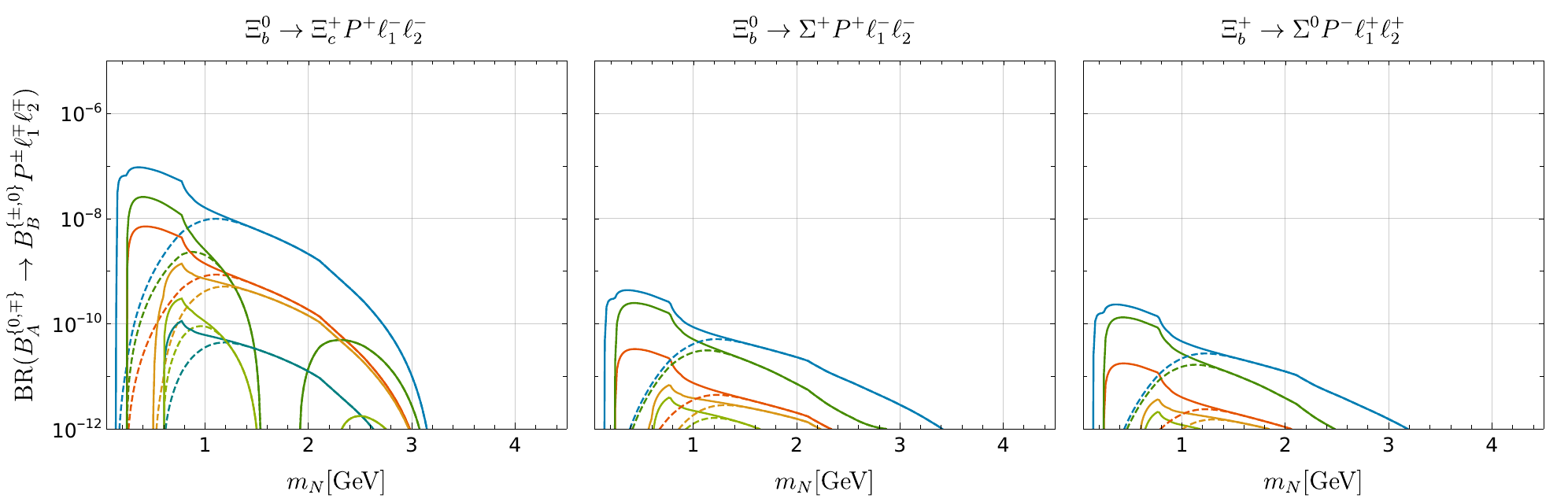}\\
        \includegraphics[scale=.5]{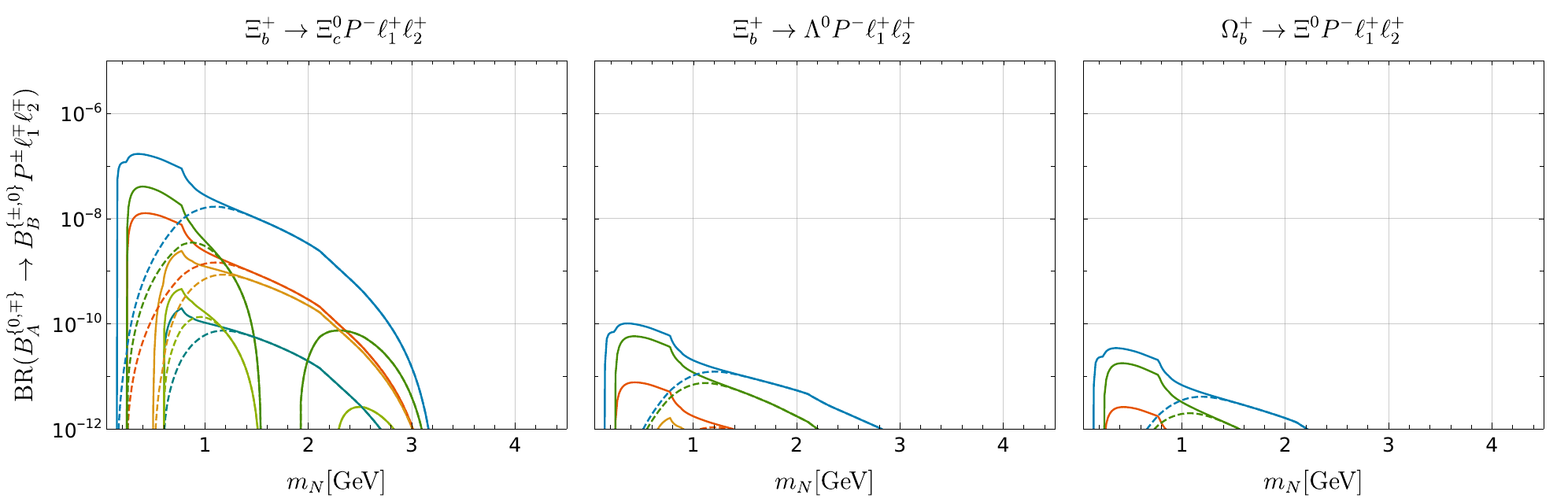}\\
        \includegraphics[scale=.5]{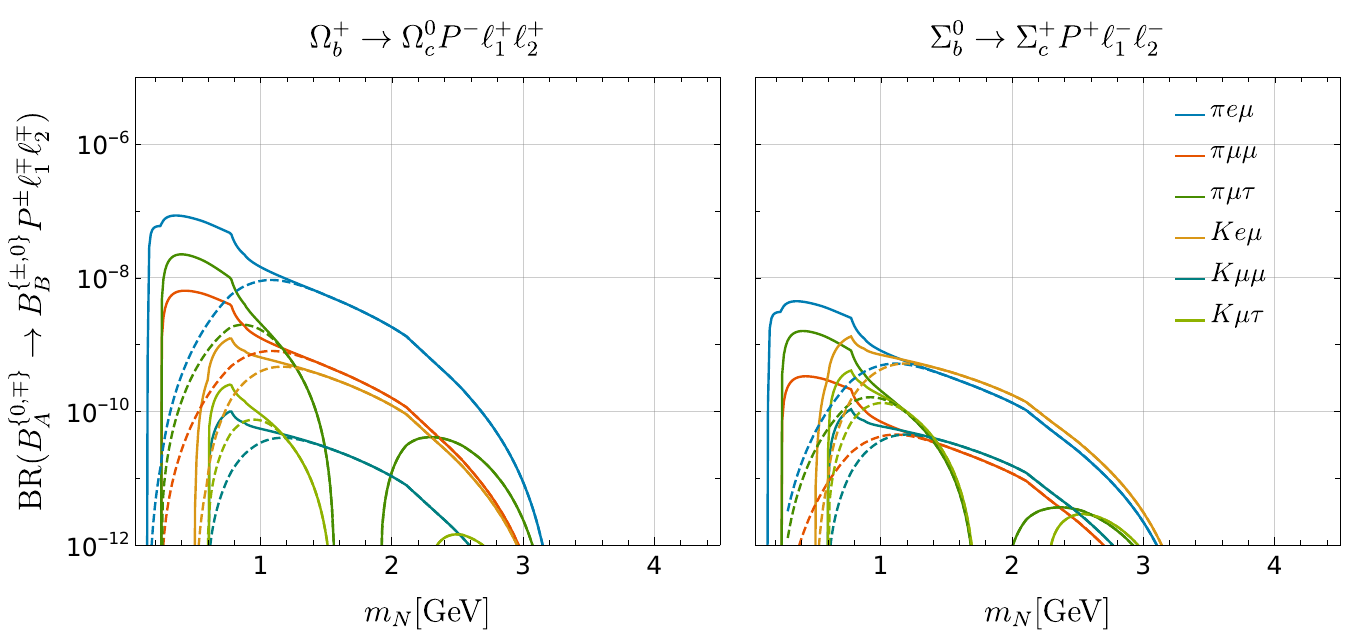}
    \end{tabular}
    \caption{Branching fraction of the $\Delta L=2$ four-body baryon decays under study as a function of the mass of the intermediate heavy neutrino, taking into account the current upper limits for the heavy-light mixings. The dashed lines are obtained by including the finite detector effect, as discussed in the text.}
    \label{fig:BrLNV}
\end{figure*}

In addition, we have computed the branching fraction of the LNV ($\Delta L=2$) four-body baryon decays under study, as a function of the intermediate heavy neutrino mass. We use the current upper limits for the heavy-light mixing parameters
from Ref.\cite{Fernandez-Martinez:2016lgt}, which are consistent, or of the same order, with those from Ref. \cite{Abada:2022wvh}
\eq{
|U_{eN}|\le& 0.005, |U_{\mu N}|\le 0.021, |U_{\tau N}|\le 0.075, 
}
where we have taken these parameters as constants for the numerical evaluation. This approach gives a generic estimate of the branching ratio. A refined analysis of particular cases would require the incorporation of the specific mass or process dependence, considered for the extraction of the bounds.
The results are shown in Fig.~\ref{fig:BrLNV}, where the solid lines are the Branching ratios without including the detector effect, and the dashed lines are the corresponding ones when it is included.

The other baryon transitions are less optimistic than the $\Lambda_b$ decay. However, some of the expected branching fractions are of $\mathcal{O}(10^{-7})$, which is not far from the expected sensitivities at the HL-LHC experiment. Thus, these transitions may be potential complementary sources for constraining the parameter space of mass and mixing, for neutrino masses around $1-3$ GeV.

\section{Conclusions} \label{sec:conclusions}

We have studied the four-body decay of heavy baryons with two charged leptons in the final state, focusing on both LNC and LNV processes, mediated by hypothetical new heavy neutrino states. Our analysis extends beyond the simplified assumption of a single heavy neutrino mixing with the active sector, by including two nearly degenerate heavy neutrinos. We profit from the so-called recast function, to illustrate the effect of the two-almost degenerate heavy neutrinos, as compared to the single case, on the exclusion region for the heavy-light mixings, for different degeneracy levels and relative phases.\\
A large set of heavy hadron transitions was considered, incorporating the relevant hadronic form factors, and all the kinematically allowed lepton pair possibilities (except for the electron pair production, due to the strong constraints in the associated heavy-light mixing).
We determined the potential exclusion region for the mass and heavy-light mixing parameters of the neutrinos driving the decay. A comparison with the bounds obtained from tau and meson decays was performed, whenever possible, exhibiting that these kind of hadronic decays may be useful to provide complementary information. The same observation is made when comparing with bounds from direct searches. The corresponding branching ratios were computed for the current upper limit of the heavy-light mixings as a function of the heavy neutrino mass.
We exhibited the finite-size detector effects, which became relevant in the low mass region of the heavy neutrino.\\
Particular attention has been paid to LNV channels with different-flavor leptons, as they may provide a distinctive opportunity to probe interference patterns, arising from lepton exchange. However, such interference turned out to be negligible, and we found that similar observations have been made in three and four body meson decays \cite{Cvetic:2013eza,Cvetic:2014nla}.
Our results complement previous works, providing possible new constraints on the parameters of the new heavy states responsible for these transitions. We remark on the role of the heavy baryon sector as part of the current search for new heavy resonant neutrino effects, especially in processes involving tau leptons, where experimental constraints remain less restrictive compared to the electron and muon sectors.

\appendix

\section{Weak transitions form factors} \label{ap:ff}
Here, we summarize the form factors used in the calculation of the baryon decays, considering the light-front  quark model. The form factors $f_3$ and $g_3$ are not obtained, due to the method used, and set to zero, which is valid in the heavy quark limit.
\subsection{\texorpdfstring{$\Xi_b\to\Xi_c$}{Xib-to-Xic}}

For this transition, we rely on  Ref.~\cite{Ke:2024aux}. The form factors $f_i$ and $g_i$ in Eq. (\ref{ff}) are written as 
\begin{align}
    f_i&= {\rm cos}^2\theta f_i^s + {\rm sin}^2\theta f_i^v\,,\nonumber\\
    g_i&= {\rm cos}^2\theta g_i^s + {\rm sin}^2\theta g_i^v\,,
\end{align}
where the superscripts $s$ and $v$ refer to transitions between particles with isospin $0$ and $1$, respectively.
The mixing angle was fitted to be about $16.27^\circ\pm 2.30^\circ$ or $85.54^\circ\pm 2.30^\circ$ (they use $16.27^\circ$ throughout their paper).\\
The energy dependence of $f_i^s,g_i^s,f_i^v$ and $g_i^v$ ($i=1,2$) form factors are parameterized, in a polynomial form, by:
\begin{equation}\label{eq:2pff}
    F(q^2)= F(0)\left[ 1+a \left(\frac{q^2}{M_{B_i}^2} \right)+b \left( \frac{q^2}{M_{B_i}^2} \right)^2 \right]\,,
\end{equation}
where $F(0)$ is the form factor at $q^2=0$, $a$ and $b$ are two parameters to be fitted from numerical results and $M_{B_i}$ is the mass of the initial baryon. The numerical values of the parameters $F(0)$, $a$ and $b$ are given in Table \ref{tab:xi}. 

\begin{table}[htb]
    \centering
    \begin{tabular}{cccc}
    \hline
    \hline
       F & F($0$) & $a$ & $b$  \\
    \hline 
    $f_1^s$ & 0.467 & 2.19 & 2.09 \\
    $f_2^s$ & -0.185 & 2.70 & 3.01\\
    $g_1^s$ & 0.448 & 2.09 & 1.92 \\
    $g_2^s$ & -0.052 & 3.32 & 4.54\\
    $f_1^v$ & 0.471 & 2.57 & 2.40\\
    $f_2^v$ & 0.378 & 2.56 & 2.81\\
    $g_1^v$ & -0.149 & 2.07 & 2.00\\
    $g_2^v$ & -0.006 & 2.95 & 2.98\\
    \hline
    \hline
    \end{tabular}
    \caption{Parameters on the form factor Eq.~(\ref{eq:2pff}), for the transition $\Xi_b\to\Xi_c$. Taken from Ref.~\cite{Ke:2024aux}.}
    \label{tab:xi}
\end{table}

\subsection{\texorpdfstring{$\Sigma_b\to\Sigma_c$ and $\Lambda_b\to\Lambda_c$}{Sb-to-Sc, and Lb-to-Lc}} 

For these transitions, we rely on Ref.~\cite{Ke:2019smy}. A pole form was employed to parameterize the energy dependence of the form factors $f_i$ and $g_i$ in Eq. (\ref{ff})
\begin{equation} \label{eq:3pff}
    F(q^2) = \frac{F(0)}{\left( 1-\frac{q^2}{M_{B_i}^2}\right) \left[ 
 1-a \left( \frac{q^2}{M_{B_i}^2}\right) + b \left( \frac{q^2}{M_{B_i}^2}\right)^2\right]}\,,
\end{equation}
where $M_{B_i}$ is the mass of the initial baryon. 
The numerical values of $F(0)$, $a$ and $b$ are given in Tables \ref{tab:sigma} and \ref{tab:lambda} for the $\Sigma_b\to\Sigma_c$ and $\Lambda_b\to\Lambda_c$, respectively.
\begin{table}[htb]
    \centering
    \begin{tabular}{cccc}
    \hline
    \hline
       F & F($0$) & $a$ & $b$  \\
    \hline 
    $f_1$ & 0.494 & 1.73 & 1.40 \\
    $f_2$ & 0.407 & 1.03 & 0.830 \\
    $g_1$ & -0.156 &  1.03 & 0.355 \\
    $g_2$ & -0.0529 & 1.58 & 2.74 \\
    \hline
    \hline
    \end{tabular}
    \caption{Parameters of the $\Sigma_b\to\Sigma_c$ form factors  (Eq. (\ref{eq:3pff})). Taken from Ref.~\cite{Ke:2019smy}.}
    \label{tab:sigma}
\end{table}
\begin{table}[t]
    \centering
    \begin{tabular}{cccc}
    \hline
    \hline
       F & F($0$) & $a$ & $b$  \\
    \hline 
    $f_1$ & 0.488 & 1.04 & 0.38 \\
    $f_2$ & -0.180 & 1.71 & 0.58 \\
    $g_1$ & 0.470 &  0.953 & 0.361 \\
    $g_2$ & -0.0479 & 2.06 & 0.89 \\
    \hline
    \hline
    \end{tabular}
    \caption{
    Parameters of the $\Lambda_b\to\Lambda_c$ form factors  (Eq. (\ref{eq:3pff})). Taken from Ref.~\cite{Ke:2019smy}.}
    \label{tab:lambda}
\end{table}

\subsection{Remaining weak transitions}

For the following transitions, we rely on Ref. \cite{Zhao:2018zcb}:
\begin{align}
    &\Lambda_b^0 (bud) \to p(uud)\,,\nonumber\\
    &\Xi_b^0(bus)\to\Sigma^+(uus)\,\nonumber,\\
    &\Xi_b^-(bds)\to\Sigma^0(uds)/\Lambda(uds)\,,\nonumber\\
    &\Omega_b^-(bss)\to\Xi^0(uss)/\Omega_c^0(css)\,.\nonumber
\end{align}
The following parameterization was used to describe the $q^2$ distributions of the form factors in Eq. (\ref{ff}):
\begin{equation} 
    F(q^2)=\frac{F(0)}{1- \frac{q^2}{m_{\rm fit}^2}+\delta \left( \frac{q^2}{m_{\rm fit}^2}\right)^2}\,,\label{eq:ffmfit}
\end{equation}
where $F(0)$ is the form factor at $q^2=0$. $m_\text{fit}$ and $\delta$ are two parameters to be fitted from numerical results. In Table \ref{tab:b-bar}, the numerical values for the parameters are collected.

\begin{table}[htb]
    \centering
    \begin{tabular}{cccc|cccc}
    \hline
    \hline
       F & F($0$) & $m_\text{fit}$ & $\delta$  &
       F & F($0$) & $m_\text{fit}$ & $\delta$\\
    \hline 
    $f_1^{\Lambda^0_b\to p}$ & 0.282 & 4.66 & 0.30 & $f_2^{\Lambda^0_b\to p}$ & -0.084 & 3.94 & 0.37\\
    $g_1^{\Lambda^0_b\to p}$ & 0.273 & 4.81 & 0.32 & $g_2^{\Lambda^0_b\to p}$ & -0.012 & 3.67 & 0.37\\
    \hline
    $f_1^{\Xi^0_b\to\Sigma^+}$ & 0.260 & 4.46 & 0.34 &$f_2^{\Xi^0_b\to\Sigma^+}$ & -0.086 & 3.84 & 0.40\\
    $g_1^{\Xi^0_b\to\Sigma^+}$ & 0.251 & 4.60 & 0.36 &
    $g_2^{\Xi^0_b\to\Sigma^+}$ & -0.012 & 3.56 & 0.41 \\
    \hline
    $f_1^{\Xi^-_b\to\Sigma^0}$ & 0.260 & 4.46 & 0.34 &
    $f_2^{\Xi^-_b\to\Sigma^0}$ & -0.086 & 3.84 & 0.40 \\
    $g_1^{\Xi^-_b\to\Sigma^0}$ & 0.251 & 4.60 & 0.36 &
    $g_2^{\Xi^-_b\to\Sigma^0}$ & -0.012 & 3.56 & 0.41 \\
    \hline
    $f_1^{\Xi_b^-\to\Lambda}$ & 0.260 & 4.46 & 0.34 &
    $f_2^{\Xi_b^-\to\Lambda}$ & -0.086 & 3.84 & 0.40 \\
    $g_1^{\Xi_b^-\to\Lambda}$ & 0.251 & 4.60 & 0.36 &
    $g_2^{\Xi_b^-\to\Lambda}$ & -0.012 & 3.56 & 0.41 \\
    \hline
    $f_1^{\Omega_b^-\to\Xi^0}$ & 0.169 & 3.30 & 0.64 &
    $f_2^{\Omega_b^-\to\Xi^0}$ & 0.193 & 3.45 & 0.49 \\
    $g_1^{\Omega_b^-\to\Xi^0}$ & -0.033 & 4.38 & 0.20 &
    $g_2^{\Omega_b^-\to\Xi^0}$ & -0.041 & 4.32 & 0.65 \\
    \hline
    $f_1^{\Omega_b^-\to\Omega_c^0}$ & 0.566 & 3.92 & 0.49 &
    $f_2^{\Omega_b^-\to\Omega_c^0}$ & 0.531 & 4.08 & 0.41 \\
    $g_1^{\Omega_b^-\to\Omega_c^0}$ & -0.170 & 4.80 & 0.23 &
    $g_2^{\Omega_b^-\to\Omega_c^0}$ & -0.031 & 9.02 & 5.05 \\
    \hline
    \hline
    \end{tabular}
    \caption{Parameters of the form factors for bottom baryon decays in Eq.~(\ref{eq:ffmfit}). Taken from Ref.~\cite{Zhao:2018zcb}.} 
    \label{tab:b-bar}
\end{table}

The physical transition form factors should be multiplied by the corresponding overlap factor (overlap of wave functions in the initial and final states) given in Table \ref{tab:overlap}.
\begin{table}[htb]
    \centering
    \begin{tabular}{cc}
    \hline
    \hline
       Transition & Overlap Factors   \\
    \hline 
    $\Lambda_b^0(bud)\to p(uud)$ & $\frac{1}{\sqrt{2}}$\\
    $\Xi_b^0(bus)\to\Sigma^+(uus)$ & $\frac{1}{\sqrt{2}}$ \\
    $\Xi_b^-(bds)\to\Sigma^0(uds), \,\Lambda(uds)$ & $\frac{1}{2},\, -\frac{1}{2\sqrt{3}}$ \\
    $\Omega_b^-(bss)\to\Xi^0(uss),\, \Omega_c^0(css)$ & $-\frac{1}{\sqrt{3}},\, 1$ \\
    \hline
    \hline
    \end{tabular}
    \caption{Overlap factors for the baryon transitions.}
    \label{tab:overlap}
\end{table}

\section{Interferences} \label{ap:recast}

The computation of the different LNC and LNV processes can be carried out in the standard form, by squaring the corresponding amplitudes. Here, we present the relations between the different $P_{1j}$ (or $P_{2j}$)  terms, Eq.~(\ref{eq:P1i}),  to recast them, as discussed in the body of this manuscript, in the narrow width approximation:
\begin{itemize}
    \item Same channel and same heavy neutrino
\end{itemize}
\begin{equation}
    P_{1j} P^*_{1j}=\frac{\pi}{m_N \Gamma_N} \delta \left(a_1^2-m_N^2 \right),
    \label{eq:B1}
\end{equation}
where $\Gamma_{N}=\Gamma_1\approx\Gamma_2$.
\begin{itemize}
    \item Interference of same channel but different heavy neutrino
\end{itemize}
\begin{equation}
    P_{14} P^*_{15} =
    \kappa(y)(1+iy)
   \frac{\pi}{m_N \Gamma_N}\delta \left(a_1^2-m_N^2 \right),
\end{equation}
where $y\equiv\Delta m_N /\Gamma_{N}$ and $\kappa(y)=1/(1+y^2)$. Thus, these two equations are not other but the known results for a single channel \cite{Cvetic:2023bwr,Abada:2022wvh}.\\

In the case of LNV there are two diagrams associated with the leptons interchange, introducing a $P_{2j}$ term. The square amplitudes follow the same structure as given above, with the corresponding $a_1^2$ and $a_2^2$ momenta in the delta function. The remaining relations are:
\begin{itemize}
    \item Interference of different channels but the same neutrino
\end{itemize} 
\begin{align}
    {\rm Re}(P_{1j} P^*_{2j}) =
    \frac{\pi}{m_N\Gamma_N} \delta \left(a_1^2-m_N^2 \right) \delta \left(a_2^2-a_1^2  \right).
\end{align} 

\begin{align}
    {\rm Im}(P_{1j} P^*_{2j}) =
    \pi\left(\frac{\delta (a_1^2-m_N^2)}{a_2^2-m_N^2} + \frac{\delta (a_2^2-m_N^2)}{a_1^2-m_N^2}  \right).
\end{align} 
This equation is valid for $a_1^2\neq a_2^2$ and null otherwise. We recover the single channel result, Eq.~(\ref{eq:B1}), for $a_1^2= a_2^2$.
\begin{itemize}
    \item Interference of different channels and different neutrino
\end{itemize}
\begin{align}
    P_{1i} P^*_{2j} =
    \frac{i \pi}{2} \kappa(y)(1+iy)&\Big( \delta (a_2^2-m_N^2)P_{1N}\nonumber\\
    &- \delta (a_1^2-m_N^2) P_{2N}^* \Big)\,.
\end{align} 
They have similar expressions as in the single channel, but with the correction from the second channel contribution.
Notice that they recover the single channel result, when $a_1$ and $a_2$ are the same.

\begin{acknowledgments}
The work of F.F. is funded by \textit{Estancias Posdoctorales por México, Estancia Posdoctoral Iniciales, Conahcyt-Secihti}. We acknowledge the support of CONAHCYT-SECIHTI, México, and the support of DGAPA-PAPIIT UNAM, under Grant No. IN110622. D.P.S. thanks CONAHCYT-SECIHTI for the financial support during his Ph.D. studies.
\end{acknowledgments}

\bibliographystyle{aipnum4-2.bst}
\bibliography{Lambda_b}% Produces the bibliography via BibTeX.

\end{document}